\documentclass[conference]{IEEEtran}

\usepackage{balance}

\usepackage{listings}
\usepackage{hyperref}
\usepackage{hyperref}
\hypersetup{
 unicode=true,
 pdfstartview={FitH},
 colorlinks=true,
 citecolor=blue,
 linkcolor=blue,
 urlcolor=blue
}

\renewcommand{\footnoterule}{%
	\kern -3pt
	\hrule% width \textwidth height 1pt
	\kern 2pt
}

\usepackage{enumitem}
\usepackage{cite}
\usepackage{amsmath,amssymb,amsfonts}
\usepackage{algorithmic}
\usepackage{graphicx}
\usepackage{textcomp}
\usepackage{xcolor}
\usepackage[a4paper, total={184mm,239mm}]{geometry}
\def\BibTeX{{\rm B\kern-.05em{\sc i\kern-.025em b}\kern-.08em
    T\kern-.1667em\lower.7ex\hbox{E}\kern-.125emX}}

\usepackage[absolute]{textpos}
\newcommand{\myboxwidth}{100}
\newcommand{\myboxheight}{13}
\newcommand{\myboxposx}{55}
\newcommand{\myboxposy}{270}
\usepackage[skins]{tcolorbox}

\begin{document}

\bstctlcite{IEEEexample:BSTcontrol}

\title{EVEREST: A design environment for extreme-scale big data analytics on heterogeneous platforms}

\author{\IEEEauthorblockN{Christian Pilato\IEEEauthorrefmark{2}, Stanislav Bohm\IEEEauthorrefmark{6}, Fabien Brocheton\IEEEauthorrefmark{9}, Jeronimo Castrillon\IEEEauthorrefmark{4}, Riccardo Cevasco\IEEEauthorrefmark{8},\\ Vojtech Cima\IEEEauthorrefmark{6}, Radim Cmar\IEEEauthorrefmark{10},
    Dionysios Diamantopoulos\IEEEauthorrefmark{1}, Fabrizio Ferrandi\IEEEauthorrefmark{2}, 
    Jan Martinovic\IEEEauthorrefmark{6}, \\ Gianluca Palermo\IEEEauthorrefmark{2}, Michele Paolino\IEEEauthorrefmark{7},
    Antonio Parodi\IEEEauthorrefmark{5}, Lorenzo Pittaluga\IEEEauthorrefmark{8}, Daniel Raho\IEEEauthorrefmark{7},\\ Francesco Regazzoni\IEEEauthorrefmark{3}, Katerina Slaninova\IEEEauthorrefmark{6}, Christoph Hagleitner\IEEEauthorrefmark{1}}\\[-0.3cm]
\IEEEauthorblockA{
\textit{\IEEEauthorrefmark{1}IBM Research Europe, Switzerland},
\textit{\IEEEauthorrefmark{2}Politecnico di Milano, Italy},
\textit{\IEEEauthorrefmark{3}Universit\`{a} della Svizzera italiana, Switzerland},\\
\textit{\IEEEauthorrefmark{4}Technische Universit\"{a}t Dresden, Germany},
\textit{\IEEEauthorrefmark{5}Centro Internazionale di Monitoraggio Ambientale, Italy},\\
\textit{\IEEEauthorrefmark{6}IT4Innovations, VSB – Technical University of Ostrava, Czech Republic}, \textit{\IEEEauthorrefmark{7}Virtual Open System, France},\\
\textit{\IEEEauthorrefmark{8}Duferco Energia, Italy},
\textit{\IEEEauthorrefmark{9}NUMTECH, France},
\textit{\IEEEauthorrefmark{10}Sygic, Slovakia}
}\vspace{-16pt}
}

\maketitle

\begin{abstract}
High-Performance Big Data Analytics (HPDA) applications are characterized by huge volumes of distributed and heterogeneous data that require efficient computation for knowledge extraction and decision making. Designers are moving towards a tight integration of computing systems combining HPC, Cloud, and IoT solutions with artificial intelligence (AI). Matching the application and data requirements with the characteristics of the underlying hardware is a key element to improve the predictions thanks to high performance and better use of resources.

We present EVEREST, a novel H2020 project started on October 1st, 2020 that aims at developing a holistic environment for the co-design of HPDA applications on heterogeneous, distributed, and secure platforms. EVEREST focuses on programmability issues through a {\em data-driven} design approach, the use of hardware-accelerated AI, and an efficient runtime monitoring with virtualization support. In the different stages, EVEREST combines state-of-the-art programming models, emerging communication standards, and novel domain-specific extensions. We describe the EVEREST approach and the use cases that drive our research. 
\end{abstract}

\begin{textblock}{\myboxwidth}(\myboxposx,\myboxposy)
  \noindent%
  \begin{tcolorbox}[left skip=0pt,width={\dimexpr\myboxwidth mm},boxrule=1pt,nobeforeafter,enhanced jigsaw,height={\dimexpr\myboxheight mm},colback=white]
\begin{center}
	\footnotesize Paper accepted for presentation at the IEEE/EDAC/ACM Design, Automation and Test in Europe Conference and Exhibition (DATE 2021)
\end{center}  
  \end{tcolorbox}
\end{textblock}

\section{Introduction}

Thanks to the pervasive use of technology, we collect volumes of data that are doubling every two years. {\em Big Data analytics} aims at extracting hidden knowledge from the data and take valuable actions, often with the support of {\em artificial intelligence} (AI)~\cite{VENIERIS18}. For example, precision medicine can create custom medical practices after analyzing patients' data that are continuously collected. Data sources are inherently distributed and heterogeneous, while accurate predictions and decisions often lead to demanding processing requirements. High-Performance Big Data Analytics (HPDA) applications expose a high parallelism and can benefit from hardware acceleration, but the limited bandwidth and the excessive power consumption for data movements put high pressure on communication and storage. HPDA applications thus require (1) efficient hardware acceleration for data processing~\cite{RDF15,FPGA17}, (2) communication cost reduction, by moving the computation closer to the data sources~\cite{KAMBATLA20142561,MINUTOLI16}, and (3) data protection from unauthorized accesses during all application phases~\cite{jin2020security}.
To accelerate data processing, HPDA systems will combine heterogeneous nodes~\cite{9256819}, with general-purpose processors and FPGA devices, across different technologies (e.g., HPC, cloud computing, and edge devices). 
Optimizing communication and storage requires to match the characteristics of the target system (e.g., distribution of the nodes and communication infrastructure, size of on-chip and off-chip memories, and number of memory channels) and the applications (e.g., data distribution and access patterns). AI methods and security threats may impose additional application and architectural constraints, especially when the data and the computation are geographically distributed. 
Since a one-fits-all solution is impossible, future HPDA systems will be {\em data-driven} with application-specific optimizations to match the application requirements, the nature and locality of the data, and the hardware characteristics~\cite{KAMBATLA20142561}. Programming such systems necessitates the use of complex data management techniques and domain-specific annotations, which are not well supported in current design frameworks. This leaves most of the effort to the application developers. Solutions to these programmability issues demand methods to represent functional and non-functional properties, drive the hardware-software compilation, and dynamically manage the underlying distributed hardware in order to obtain fast, scalable, and secure HPDA systems. 

The EU project EVEREST (dEsign enVironmEnt foR Extreme-Scale big data analyTics on heterogeneous platforms - \url{http://www.everest-h2020.eu})  proposes a design environment for HPDA applications on distributed and heterogeneous systems. The EVEREST target system seamlessly combines nodes with IBM POWER9 CPUs and coherent FPGA accelerators (for cloud computing), and disaggregated FPGA devices~\cite{8071053} (for edge computing). The EVEREST design environment complements state-of-the-art programming models (e.g., OpenCL, SYCL, OpenMP) with domain-specific extensions to (1) provide extra characteristics of the algorithms and data, (2) exploit the available hardware resources with alternative code/hardware variants, (3) promote the use of high-level synthesis (HLS)~\cite{TCAD16} for generating AI accelerators, and (4) improve the dynamic control of the distributed execution~\cite{cima2018hyperloom,Gadioli19Margot}. 
 
\section{EVEREST Approach}

Our {\bf EVEREST System Development Kit (SDK)} is a design environment to ease the description, optimization and execution of Big Data applications with heterogeneous data sources onto FPGA-based architectures, operating at design and run time.

At {\em design time}, we focus on (1) the application description along with non-functional requirements, (2) the generation of several hardware and software variants, and (3) the customization of the distributed memory architecture. We aim at developing a data-driven hardware/software compilation framework that takes as input an application description using a combination of workflow libraries, AI libraries and frameworks, and domain-specific extensions. The compilation engine explores code variants and uses HLS for generating hardware accelerators. We represent the resulting application with mainstream parallel programming models (like SYCL). Flexible memory managers will enable to co-optimize computation, communication, and storage, to move the computation closer to the data, and to implement hardware-assisted data protection. 

At {\em runtime}, we build a virtualized environment to dynamically select the code variant to execute for each task, based on the workload and data conditions. The virtualized environment will abstract hardware characteristics of the EVEREST nodes (based on different CPU architectures e.g., x86 on the cloud and ARM/RISC-V on the edge) to present an integrated execution environment for the applications. This combined solution allows designers to match the data requests with the underlying hardware to optimize the data transfers, exploit the spatial parallelism with the hardware accelerators, and react to changes in the workload conditions.

\section{Data-driven Compilation Framework}

\subsection{Application specification and definition of requirements}\label{sec:data}

The EVEREST design framework receives as input the application description (i.e., a workflow pipeline where each node can be specified in C/C++ or with proper AI libraries). Industry-grade applications often encompass end-to-end data processing workflows composed of a large number of interconnected computational tasks of various granularity. EVEREST will feature a scalable platform based on HyperLoom~\cite{cima2018hyperloom} for describing and executing complex workflows in large scale distributed environments with various virtualized heterogeneous resources. The envisioned platform aims to improve resource utilization and reduces the overall workflow processing time.

Application experts are offered {\bf embedded domain-specific languages} (DSLs) to express the semantics and security requirements of computational tasks to enable high-level code optimizations. 
DSL extensions have been successfully demonstrated in many domains, such as  computational fluid dynamics~\cite{rink_rwdsl18}, hybrid particle-mesh simulations~\cite{karol_toms18}, tensor expression optimizations~\cite{rink_gpce18,rink_array19,chen2018tvm}, and dataflow languages~\cite{ertel_cc18}. 
EVEREST proposes a data-centric approach for security, dealing with confidentiality, authentication and integrity of the data handled by the system with {\bf hardware-assisted data protection} applied to both edge devices and data center nodes. 
EVEREST will propose a comprehensive library of optimized accelerators for memory and near memory encryption, fitting the area, energy and performance constraints of the platforms. We will include information flow tracking, monitoring, and protection against malicious uses, including side-channel and buffer-overflow attacks~\cite{8356053}. 

EVEREST aims at developing a unified MLIR representation for the transparent support of several high-level ML frameworks (e.g., TensorFlow or PyTorch) and high-level optimizers (e.g., XLA, Glow, TVM)~\cite{SZE17}. 

\begin{figure}[t]
    \centering
    \includegraphics[width=0.95\columnwidth]{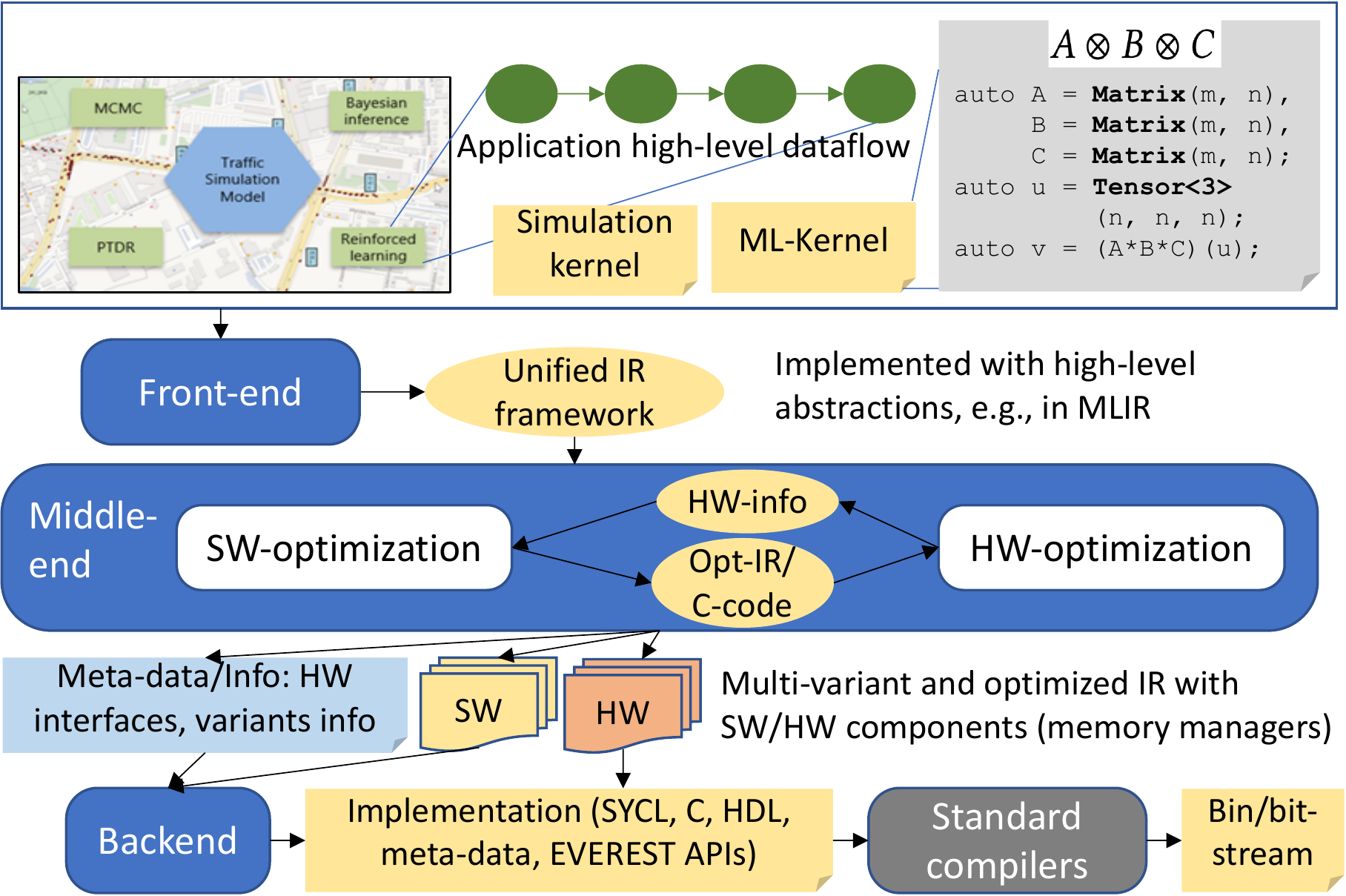}
\caption{\vspace{-6pt}Overview of the data-driven compilation flow.}\vspace{-6pt}
    \label{fig:comp-flow}
\end{figure}

\subsection{Generation of software and hardware variants}

DSLs will be used to concisely express performance-critical functionality and annotate data characteristics and requirements. Tensors and particles are two examples of EVEREST data-centric programming abstractions that will enable optimization of data communications and generation of custom memory subsystems. DSLs for expression languages will enable highly-optimized kernel generation either in software or hardware to enlarge the optimization space~\cite{rink_gpce18}, while allowing more control for provably safe execution~\cite{rink_array19}.

For a higher-level coordination of the workflow kernels, EVEREST will look at functional abstractions to implicitly express the application dataflow~\cite{ertel_cc18,ertel_haskell19} and its integration on HyperLoom~\cite{cima2018hyperloom}.
The compiler front-end unifies the orchestration and the kernel specifications into a single MLIR as shown in Fig.~\ref{fig:comp-flow}.
We will extend the LLVM compilation framework~\cite{lattner07} with dedicated MLIR dialects~\cite{mlir} for domain-specific kernels. The tool chain will support standard exchange formats used in machine learning (e.g., NNEF or ONNX). 

The middle-end of the compilation flow will rely on high-level architecture models~\cite[Chapter 6]{castrillon14_springer}\cite{ieee-2804-2019} and simulators~\cite{lowe-power_gem5_2020,menard_samos17} to explore the design space and create {\bf multiple hardware and software variants}. 
These variants are performance/energy trade-offs that are exposed to the runtime system. 
For instance, a software-only implementation could explore layouts of particles as array-of-structures or structure-of-arrays, or could tile complex tensor expressions to fit the memory hierarchy while allowing different threading implementations for the runtime. 
Hardware variants could implement a chain of tensor operations directly on the FPGA logic before writing back to main memory. 
Hardware/software partitioning will be driven by annotations and the two parts will be co-optimized, including hardware estimations for code-snippets (cf. Fig.~\ref{fig:comp-flow}).

EVEREST will leverage FPGA resources to create {\bf hardware accelerators with high-level synthesis}, especially for data-intensive and AI tasks. In EVEREST, we use Bambu, an open-source HLS tool based on both GCC and LLVM~\cite{6645550}. Bambu will optimize execution and memory bandwidth of accelerators. Data distribution introduces additional challenges in terms of variable read/write latency and energy based on the location of the data. Since the memory behavior of an application ranges from statically predictable patterns~\cite{FPGA17} to irregular memory accesses~\cite{RDF15}, we will use a {\bf fully automated and transparent memory management} at both compile time and runtime with a combination of polyhedral-based transformations~\cite{WANG14}, multi-port memories~\cite{PILATO17} and dedicated micro-architectures to schedule the memory accesses~\cite{MINUTOLI16}, interleave the memory requests and hide the communication latency with the distributed memories~\cite{PILATO11}. We will generate and optimize such accelerators based on the information extracted from the DSL annotations.
EVEREST will extend high-level synthesis for the automatic integration of security features, like application-specific dynamic information flow tracking~\cite{8114281,8356053}. We will also develop and use a library of cryptographic functions, to ensure data integrity, confidentiality, and authentication. Such cryptographic routines will match application requirements and dynamic behaviors. Dedicated hardware monitors will detect anomalies with respect to the expected data behaviors (timing patterns, access patterns, typical sizes and ranges), activating proper dynamic adaptation in the form of ``auto-protection''.

Given the set of variants, the backend will generate software implementation relying on state-of-the-art programming models (e.g. SYCL) to enable seamless integration in the tooling infrastructure. Meta-information about the variants will be provided to the runtime system to support dynamic selection. 
Finally, standard toolchains will be used to generate binaries and bitstreams for the target devices.

\section{Virtualization-based Runtime Optimization}

\begin{figure}[t]
    \centering
    \includegraphics[width=0.90\columnwidth]{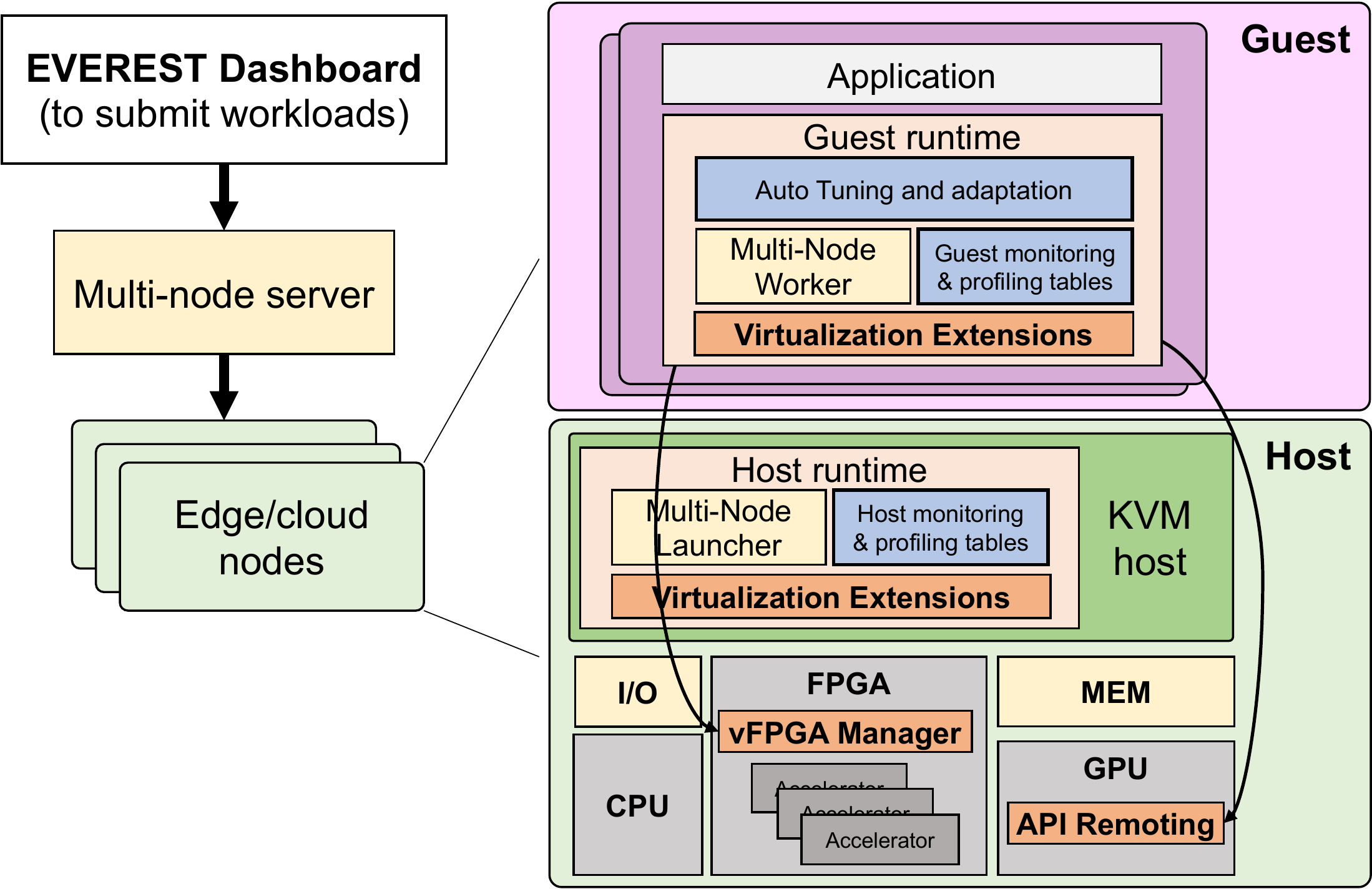}
\vspace{-8pt}\caption{Virtualized runtime environment overview.}\vspace{-12pt}
    \label{fig:virt_env}
\end{figure}

EVEREST features a {\bf distributed runtime support} to manage and coordinate the computation across the different system nodes. Tasks are defined in a way that allows runtime migration of both data and computations. 
FPGA accelerated applications and the runtime framework will be designed with a {\bf virtualized environment} to abstract the hardware resources. This approach improves efficiency and security. Also, we will be able to seamlessly move the computation between edge nodes and also between edge and cloud parts. 
The runtime layer optimizes the use of heterogeneous and distributed resources by parallel application instances running in different virtual machines (VMs). The EVEREST virtualized runtime environment automatically manage the code to run and configure the hardware based on the workload conditions and the data distribution. virtualization techniques will abstract hardware characteristics of the heterogeneous target nodes to present an integrated execution environment for the applications. As described in Section~\ref{sec:target}, the nodes may feature different CPUs (i.e., x86 in the cloud and ARM/RISC-V in the edge~\cite{sechkova2019cloud}) and accelerators (e.g., GPUs and FPGAs~\cite{ChiotakisFPGA}). Fig.~\ref{fig:virt_env} shows an overview of the EVEREST virtualized runtime environment. Its implementation includes both hypervisor and guest OS extensions to manage, optimize, and monitor the access to hardware from guest applications. These extensions provide:
\begin{enumerate}[leftmargin=1.5em]
    \item {\bf Data protection layer.} The system monitors the execution to identify malicious attacks (see Section~\ref{sec:data} and react by using the security mechanisms added by the compiler.
    \item {\bf Dynamic hardware-software adaptation strategy.} We propose an intelligent policy to select the code variant or hardware configuration to execute, among the ones pre-generated at compile time, based on the system status. 
    \item {\bf Virtualization support and hypervisor extensions.} Hardware configurable parameters, including accelerator APIs, are exposed directly to the applications inside the VMs, requiring also guest OS enhancements (e.g., drivers). 
\end{enumerate}

The EVEREST virtualization environment will interact with the underlying hardware to select the variants to execute. 
EVEREST provides {\bf dynamic application auto-tuning capabilities} based on mARGOt~\cite{Gadioli19Margot}, a dynamic decision-maker that performs an automatic selection of the variant to execute for each critical kernel identified at compile time. For example, a variant that makes heavy access to unavailable hardware resources can be replaced by a variant that fits better with the system status. This selection is based on (1) the dynamic characteristics of the target system (e.g., available resources) \cite{Paone14OCL}, (2) the optimization goal set for execution (e.g., performance or energy consumption) \cite{Gadioli18Socrates,khasanov_date20}, (3) the additional dynamic requirements (e.g., security monitoring, data features \cite{vitali19}), and (4) the available techniques for data management (e.g., data representations and distributed allocation). The selection will generalize the concept of affinity between the code variants and the available system configurations and requirements. Hardware monitors will collect the information to make the selection. 

Guest programs will configure the underlying hardware or make specific requests based on workload conditions, environment changes and the availability of specific hardware resources (e.g., communication channels, remote notes). API remoting techniques will improve data exchanges.
The distributed runtime also leverages the configuration of pre-defined hardware resources for deep learning, like reconfigurable AI networks. 

\section{EVEREST Target System}\label{sec:target}

\begin{figure}[t]
    \centering
    \includegraphics[width=\columnwidth]{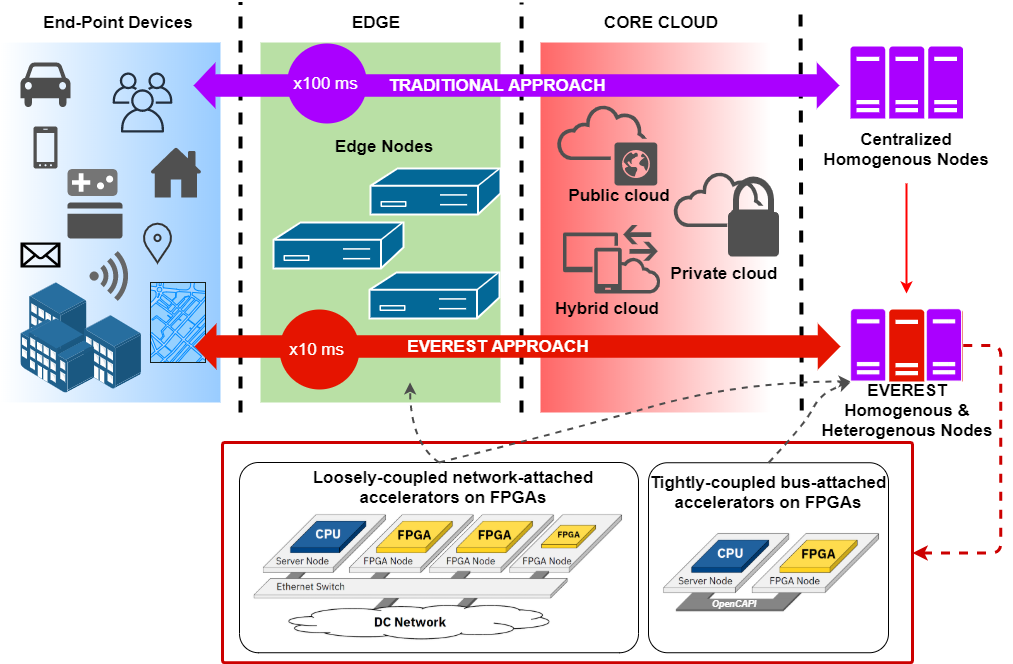}
\vspace{-18pt}\caption{EVEREST ecosystem overview.}\vspace{-10pt}
    \label{fig:ecosystem}
\end{figure}

In EVEREST, we envision a hierarchy of processing environments, as shown in Fig.~\ref{fig:ecosystem}. The outermost layer ({\em End-Point Devices}) receives the stream of data and performs initial processing under strict latency constraints and with the limited performance available in end-point devices. These requirements dictate very fast data pre-processing, inference and perhaps only limited training. Depending on the application, these edge nodes can be complemented by an inner-edge environment that does more extensive processing, training and data analysis. The inner-edge environment features more powerful hardware and less stringent requirements for real-time processing. The results of this layer are then forwarded to the core cloud services (public, private or hybrid), where more extensive analysis and model building is performed on heterogeneous hardware.

\begin{figure}[t]
    \centering
    \includegraphics[width=0.95\columnwidth]{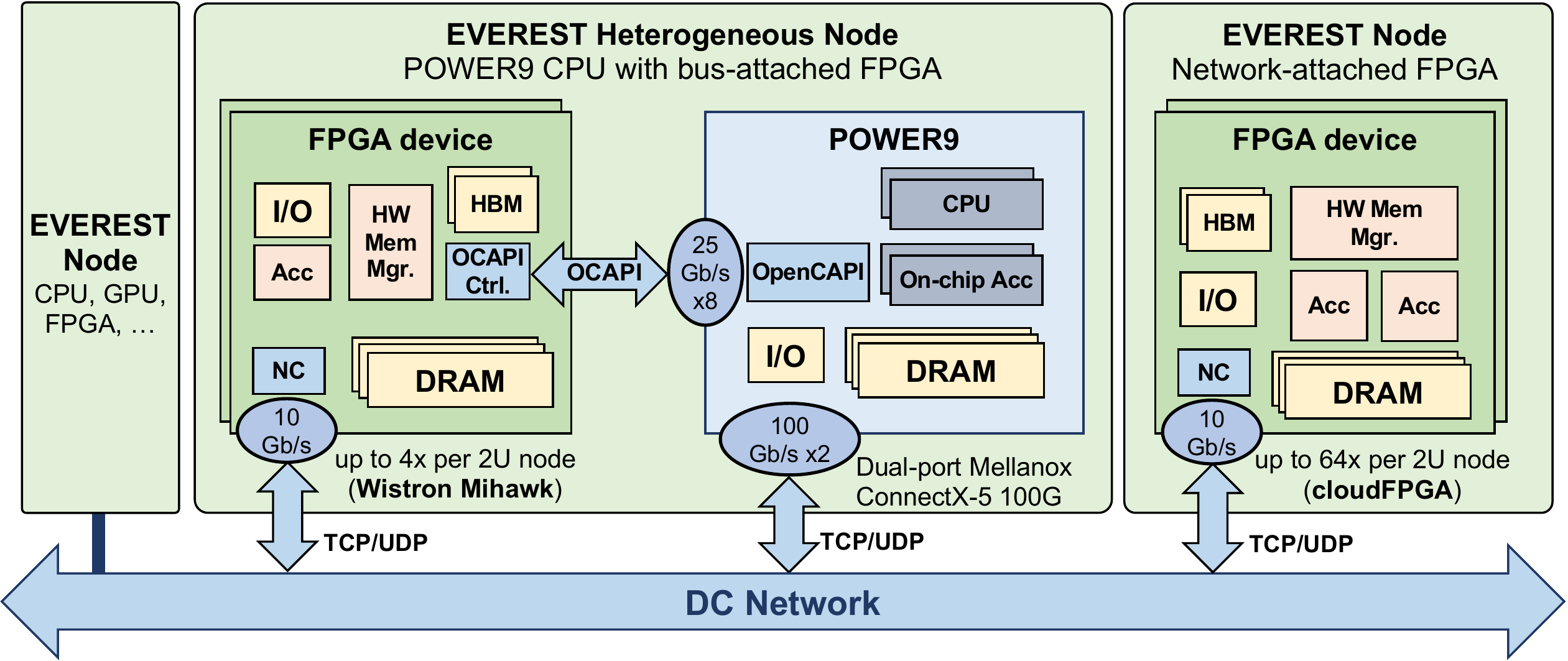}
\vspace{-6pt}\caption{EVEREST featured system as a combination of heterogeneous nodes with OpenCAPI cache coherent and TCP/UDP protocols.}\vspace{-12pt}
    \label{fig:arch}
\end{figure}

Today's edge nodes are typically scaled versions of cloud servers, which primarily combine CPUs with tightly-coupled co-processors (e.g. GPUs). However, CPUs and GPUs are optimized towards batch processing of in-memory data and can hardly provide deterministic performance for the processing of streaming data coming from the I/O channels of end-point devices. Future edge servers call for a new heterogeneous computing node tailored to the processing of streaming data at low power consumption and high energy efficiency. To support this vision, the EVEREST project targets distributed architectures composed of industry established computing nodes, with CPUs and GPUs, as well as experimental heterogeneous nodes with FPGAs. Each experimental node may feature one or more FPGA devices for hardware acceleration and one or more physical memories (either local or external to the FPGA), as shown in Fig.~\ref{fig:arch}. Such systems will run Linux as Operating System (OS) and a hypervisor to manage the hardware resources. Note that the EVEREST approach is not limited to these architectures. In fact, specifying the workflow pipelines at a higher level of abstraction, within the specifications of EVEREST SDK and virtualization technology, as discussed previously, will enable the porting of the applications to architectures with heterogeneous GPU-based nodes and end-user embedded devices.
We aim at developing a {\bf small multi-node demonstrator} based on the technology and the components available during the project’s timeline. To develop the EVEREST SDK for heterogeneous systems, we focus on two state-of-the-art FPGA-based research platforms: a CPU-managed system that rely on tightly-coupled bus-attached FPGAs \cite{did2019} and an FPGA-disaggregated system that relies on loosely-coupled network-attached FPGAs~\cite{8071053}. The final EVEREST demonstrator will feature both nodes to examine how the different architectural configurations can accommodate big data workloads at the edge and on the cloud.

In EVEREST, the POWER9 node with tightly-coupled bus-attached FPGAs forms the basic platform to research these challenges. In addition, the system is augmented with loosely-coupled network attached FPGAs to increase the parallel processing capability from multiple I/Os streams. The latter platform will be also evaluated as an edge node, where processing closer to the multiple I/O streams can offer great performance due to the low-latency and high-energy efficiency of the FPGAs. Both bus-attached FPGAs and network-attached FPGAs are envisioned as a scale-up opportunity of the POWER9 node, while multiple such nodes will be extended across the data-center, as a scale-out configuration. 

The demonstrator based on loosely-coupled network-attached FPGAs will rely on the cloudFPGA platform~\cite{8071053}. CloudFPGA is a research platform that disaggregates the FPGA accelerator from the server, turning it into a stand-alone computing resource. Such network-attached FPGAs can be deployed at large scale and independently of the number of CPU servers in the data-center (DC). The network attachment allows them to seamlessly connect with each other as well as with one or more CPUs. The resulting disaggregated heterogeneous computing infrastructure is capable to dynamically adapt to the scale of any workload. Meanwhile, large-scale applications ranging from business analytics to scientific simulations and AI have started to scale out using distributed frameworks such as Hadoop, Spark, HyperLoom, and Tensorflow.
The cloudFPGA platform enables a user to acquire, distribute, configure and operate stand-alone network-attached FPGAs at large scale in DC infrastructures. The use of a shell-role architecture combined with partial reconfiguration provides for isolation of the system management functions from the user logic within the network-attached FPGA. This approach protects the integrity of the DC network by creating a separation between privileged and non-privileged user logic functions.

\section{EVEREST Use Cases}
We drive our research with three industrial applications: (1) a weather analysis-based {\bf prediction model for the energy trading market}, (2) an application for {\bf air-quality monitoring} of industrial sites, and (3) a {\bf traffic modeling framework for intelligent transportation} in smart cities. 
The applications are representatives of future HPDA applications: they have large and heterogeneous data sets ({\em Volume} and {\em Variety}), including historical and real-time data ({\em Velocity}) with important security concerns during communication and storage ({\em Veracity}). They are also aligned with the United Nations Sustainable Development Goals (no. 7, 9, 11 and 13).

\subsection{Weather-based predictions for renewable energy production}

In 2017, for the first time, the European Union generated more electricity from wind, solar and biomass than from coal according to new analysis from \textit{Sandbag} and \textit{Agora Energiewende}. 
The European energy market shows a strong interplay between the different energy sources: for example, a drought period can affect the hydroelectric production, demanding gas generation to counterbalance. Also, the prediction of energy production from renewable sources (in particular wind) is uncertain. Renewable energy production forecasting systems currently rely on an ensemble of meteorological predictions provided by global circulation models with grid spacing between 15 and 25 km and hourly temporal resolution. This ensemble predicts variables such as 2m temperature, near-surface wind speed, incoming solar radiation, and rainfall depth, that become input of a subsequent deep learning model trying to characterize the complex input/output relationship of the given power plant under consideration. Even using ensemble approaches, large uncertainties still exist when operating forecasting systems based on meteorological variables predictions at tenths of km’s, especially when dealing with sudden local changes in cloud cover and wind intensity.

In EVEREST, we aim at reducing the cost of imbalance in case of severe meteorological ramp-up/down events. The application will forecast the energy produced by a wind farm in the next day with a 24-hour prediction on a hourly basis. 
Thanks to transparent hardware acceleration, we will be able to increase the resolution of weather forecast ensembles to better predict high-localized meteorological variations at hourly scale~\cite{lagasio2019predictive,lagasio2019synergistic}. Thanks to AI tools, we will combine the resulting weather models with historical data.  

\subsection{Air-quality monitoring in industrial sites}

Every year new publications show the impact of air quality on public health. The latest figures from the World Health Organization (WHO) show that air pollution kills an estimated seven million people worldwide every year. WHO data shows also that 9 out of 10 people breathe air containing high levels of pollutants, and that the economic impact of this pollution on health is estimated to 5.7K billions of dollars per year. Industry contributes to this impact and must adapt on the one hand to increasing regulatory constraints and on the other hand to citizen pressure, in particular due to the development of low-cost air-quality sensors providing massive amounts of (low quality) spatial information. 
To support manufacturers, NUMTECH offers Plum'air, a service that allows an industrial site to collect real-time information about the monitoring and control of the pollution. In forecast mode, it can be used as a decision tool for an industrial site to adapt its activity in order to reduce its impact, especially in the transition phase before the implementation of heavy investments in terms of emission treatment or reduction systems. In this mode, Plum'air aims at forecasting the environmental impacts due to atmospheric releases of an industrial site at local scale (within 10 km from emission sources).

In EVEREST, we forecast the environmental impacts of chemical pollutants combining high-resolution weather ensembles with local data. Together with hardware acceleration, we will be able to obtain accurate information about the environment so that the industrial site can promptly delay production activities that may have an impact (e.g., increase of atmospheric releases) or activate emission reduction treatments. 

\subsection{Traffic modeling for intelligent transportation}

Traffic modelling and prediction is a critical component for Smart Cities to build their intelligent traffic management system (ITS). Our approach for designing such a component is by creating a traffic modelling ecosystem comprised of tightly coupled processing elements such as reading big sensory data real-time and of a long-history records; traffic simulator which boosts the raw sensory data dataset into rich training sequences; traffic prediction model which learns from the training data set; route calculation as a service exploiting traffic prediction model. As the main data input into the system we will use provisioned origin-destination matrix (O/D) and a large historical data set of floating car data (FCD). FCD is represented by geo position and the speed of vehicle sensed approximately each 5 seconds from navigation devices, that is from millions of devices every day over the period of several years. However, our model will operate on selected cities (like Vienna) counting thousands of vehicles daily.
Traffic simulator simulates individual clients driving around the smart city by combining both macro and microscopic approaches, optimizing the traffic flow~\cite{Golasowski20,ptosek18,vitali19}.  The simulator calculates traffic model in near-real time while it requires access to historical records and a streaming long data chunks for ML predictions. It updates the traffic model for various conditions as well as it can generate training sequences for traffic predictions. 

In EVEREST, we will improve the key processing components of the traffic modeling eco-system. The use of efficient AI methods will allow the edge nodes to collect and process more data, while addressing all privacy and security concerns. Also in this case, the EVEREST SDK will allow non-expert designers to easily express the application requirements for the compilation framework and the runtime system.

\subsection{Why using the EVEREST SDK?}

The applications will demonstrate how the EVEREST SDK can unleash novel market opportunities for the respective companies. The EVEREST SDK will provide following benefits:
\begin{itemize}[leftmargin=1.5em]
	\item {\bf Quality of predictions}: the possibility of integrating real-time and historical data by means of AI will allow more accurate predictions. This aspect is crucial for all applications as their commercial value lies on precise and timely knowledge extracted from the data.
	\item {\bf Performance and energy efficiency}: the efficient use of heterogeneous resources and, in particular, hardware acceleration will reduce the time and the energy spent for obtaining the results with significant competitive advantage. For example, intra-day renewable energy prediction will open new market opportunities. Similarly, industrial sites require fast and efficient systems for air-quality monitoring.
	\item {\bf Dynamic adaptation}: due to the distributed and heterogeneous nature of the data (e.g., traffic data), the combination of code and hardware variants, dynamic autotuning, and virtualization will enable a transparent use of the hardware resources even in case of changes to the configurations.
	\item {\bf Design productivity}: non-expert programmers will use domain-specific extensions to express the semantics of the application and the security requirements of the data. The EVEREST SDK will automatically carry out the related optimizations, broadening the customers that can be reached with complex heterogeneous platforms.
	\item {\bf Programmability support}: the EVEREST SDK will hide the platform details to the application, enabling the porting across target platforms with different characteristics. 
\end{itemize}

\section{Concluding Remarks}   

EVEREST provides a data-driven design framework for extreme-scale Big Data applications on distributed FPGA-based architectures. The EVEREST SDK combines multiple domain-specific languages, compiler optimizations, and HLS to generate multiple code variants that are dynamically selected by matching characteristics of the application and the available hardware. Our major goals are not only to accelerate the application execution, but also to ease the design of complex AI-enabled applications by non-expert programmers, hiding most of the details of the underlaying hardware system.

\balance

\section*{Acknowledgements}

This project has received funding from the EU Horizon 2020 Programme under grant agreement No 957269. 

\bibliographystyle{IEEEtran}
\bibliography{main}

\begin{thebibliography}{10}
\providecommand{\url}[1]{#1}
\csname url@samestyle\endcsname
\providecommand{\newblock}{\relax}
\providecommand{\bibinfo}[2]{#2}
\providecommand{\BIBentrySTDinterwordspacing}{\spaceskip=0pt\relax}
\providecommand{\BIBentryALTinterwordstretchfactor}{4}
\providecommand{\BIBentryALTinterwordspacing}{\spaceskip=\fontdimen2\font plus
\BIBentryALTinterwordstretchfactor\fontdimen3\font minus
  \fontdimen4\font\relax}
\providecommand{\BIBforeignlanguage}[2]{{%
\expandafter\ifx\csname l@#1\endcsname\relax
\typeout{** WARNING: IEEEtran.bst: No hyphenation pattern has been}%
\typeout{** loaded for the language `#1'. Using the pattern for}%
\typeout{** the default language instead.}%
\else
\language=\csname l@#1\endcsname
\fi
#2}}
\providecommand{\BIBdecl}{\relax}
\BIBdecl

\bibitem{VENIERIS18}
S.~I. Venieris, A.~Kouris, and C.-S. Bouganis, ``Toolflows for mapping
  convolutional neural networks on {FPGAs}: A survey and future directions,''
  \emph{ACM Comput. Surv.}, vol.~51, no.~3, Jun. 2018.

\bibitem{RDF15}
V.~G. Castellana \emph{et~al.}, ``High level synthesis of {RDF} queries for
  graph analytics,'' in \emph{Proc. of ICCAD}, 2015, p. 323–330.

\bibitem{FPGA17}
R.~Zhao \emph{et~al.}, ``Accelerating binarized convolutional neural networks
  with software-programmable {FPGAs},'' in \emph{Proc. of FPGA}, 2017, p.
  15–24.

\bibitem{KAMBATLA20142561}
K.~Kambatla \emph{et~al.}, ``Trends in big data analytics,'' \emph{Journal of
  Parallel and Distributed Computing}, vol.~74, no.~7, pp. 2561 -- 2573, 2014.

\bibitem{MINUTOLI16}
M.~Minutoli \emph{et~al.}, ``Efficient synthesis of graph methods: A
  dynamically scheduled architecture,'' in \emph{Proc. of ICCAD}, 2016.

\bibitem{jin2020security}
C.~Jin, V.~Gohil, R.~Karri, and J.~Rajendran, ``Security of cloud fpgas: A
  survey,'' \emph{arXiv preprint arXiv: 2005.04867}, 2020.

\bibitem{9256819}
P.~{Mantovani} \emph{et~al.}, ``Agile {SoC} development with {Open ESP},'' in
  \emph{Proc. of ICCAD}, 2020, pp. 1--9.

\bibitem{8071053}
F.~{Abel} \emph{et~al.}, ``An fpga platform for hyperscalers,'' in \emph{Proc.
  of HOTI}, 2017, pp. 29--32.

\bibitem{TCAD16}
R.~Nane \emph{et~al.}, ``A survey and evaluation of {FPGA} high-level synthesis
  tools,'' \emph{IEEE Trans. CAD Integ. Cir. Sys.}, vol.~35, no.~10, Oct. 2016.

\bibitem{cima2018hyperloom}
V.~Cima \emph{et~al.}, ``Hyperloom: A platform for defining and executing
  scientific pipelines in distributed environments,'' in \emph{Proc. of
  PARMA-DITAM}, 2018, pp. 1--6.

\bibitem{Gadioli19Margot}
D.~{Gadioli}, E.~{Vitali}, G.~{Palermo}, and C.~{Silvano}, ``margot: A dynamic
  autotuning framework for self-aware approximate computing,'' \emph{IEEE
  Transactions on Computers}, vol.~68, no.~5, pp. 713--728, 2019.

\bibitem{rink_rwdsl18}
N.~A. Rink \emph{et~al.}, ``{CFDlang}: High-level code generation for
  high-order methods in fluid dynamics,'' in \emph{Proc. of RWDSL}, 2018, pp.
  1--10.

\bibitem{karol_toms18}
S.~Karol \emph{et~al.}, ``A domain-specific language and editor for parallel
  particle methods,'' \emph{ACM Trans. on Mathematical Software}, vol.~44,
  no.~3, 2018.

\bibitem{rink_gpce18}
A.~Susungi \emph{et~al.}, ``Meta-programming for cross-domain tensor
  optimizations,'' in \emph{Proc. of GPCE}, 2018, pp. 79--92.

\bibitem{rink_array19}
N.~A. Rink and J.~Castrillon, ``{TeIL}: a type-safe imperative {Tensor
  Intermediate Language},'' in \emph{Proc. of ARRAY}, 2019, pp. 57--68.

\bibitem{chen2018tvm}
T.~Chen \emph{et~al.}, ``{TVM}: An automated end-to-end optimizing compiler for
  deep learning,'' in \emph{Proc. of OSDI}, 2018, pp. 578--594.

\bibitem{ertel_cc18}
S.~Ertel, A.~Goens, J.~Adam, and J.~Castrillon, ``Compiling for concise code
  and efficient {I/O},'' in \emph{Proc. of CC}, 2018, pp. 104--115.

\bibitem{8356053}
C.~{Pilato}, K.~{Wu}, S.~{Garg}, R.~{Karri}, and F.~{Regazzoni}, ``{TaintHLS}:
  High-level synthesis for dynamic information flow tracking,'' \emph{IEEE
  Trans. CAD Integ. Cir. Sys.}, vol.~38, no.~5, pp. 798--808, 2019.

\bibitem{SZE17}
V.~{Sze} \emph{et~al.}, ``Efficient processing of deep neural networks: A
  tutorial and survey,'' \emph{Proceedings of the IEEE}, vol. 105, no.~12,
  2017.

\bibitem{ertel_haskell19}
S.~Ertel \emph{et~al.}, ``{STCLang}: State thread composition as a foundation
  for monadic dataflow parallelism,'' in \emph{Proc. of Haskell Symposium},
  2019.

\bibitem{lattner07}
C.~Lattner \emph{et~al.}, ``{Making Context-Sensitive Points-to Analysis with
  Heap Cloning Practical For The Real World},'' in \emph{Proc. of PLDI}, 2007.

\bibitem{mlir}
C.~Lattner \emph{et~al.}, ``{MLIR}: A compiler infrastructure for the end of
  {M}oore's law,'' \emph{arXiv preprint arXiv:2002.11054}, 2020.

\bibitem{castrillon14_springer}
J.~Castrillon and R.~Leupers, \emph{Programming Heterogeneous {MPSoCs}: Tool
  Flows to Close the Software Productivity Gap}.\hskip 1em plus 0.5em minus
  0.4em\relax Springer, 2014.

\bibitem{ieee-2804-2019}
{C/DA - Design Automation}, ``{IEEE} 2804-2019 - {IEEE} standard for
  software-hardware interface for multi-many-core,'' Jan. 2020.

\bibitem{lowe-power_gem5_2020}
{Lowe-Power et al.}, ``The gem5 simulator: Version 20.0+,'' \emph{arXiv
  preprint arXiv: 2007.03152}, 2020.

\bibitem{menard_samos17}
C.~Menard \emph{et~al.}, ``System simulation with gem5 and systemc: The
  keystone for full interoperability,'' in \emph{Proc. of SAMOS}, 2017, pp.
  62--69.

\bibitem{6645550}
C.~{Pilato} and F.~{Ferrandi}, ``Bambu: A modular framework for the high level
  synthesis of memory-intensive applications,'' in \emph{Proc. of FPL}, 2013.

\bibitem{WANG14}
Y.~Wang, P.~Li, and J.~Cong, ``Theory and algorithm for generalized memory
  partitioning in high-level synthesis,'' in \emph{Proc. of FPGA}, 2014.

\bibitem{PILATO17}
C.~Pilato \emph{et~al.}, ``System-level optimization of accelerator local
  memory for heterogeneous systems-on-chip,'' \emph{IEEE Trans. CAD Integ. Cir.
  Sys.}, vol.~36, no.~3, p. 435–448, Mar. 2017.

\bibitem{PILATO11}
C.~Pilato \emph{et~al.}, ``A runtime adaptive controller for supporting
  hardware components with variable latency,'' in \emph{Proc. of AHS}, 2011,
  pp. 153--160.

\bibitem{8114281}
C.~{Pilato} \emph{et~al.}, ``Securing hardware accelerators: A new challenge
  for high-level synthesis,'' \emph{IEEE Embedded Systems Letters}, vol.~10,
  no.~3, pp. 77--80, 2018.

\bibitem{sechkova2019cloud}
T.~Sechkova, E.~Barberis, and M.~Paolino, ``Cloud \& edge trusted virtualized
  infrastructure manager (vim)-security and trust in openstack,'' in
  \emph{Proc. of WCNCW}, 2019, pp. 1--6.

\bibitem{ChiotakisFPGA}
S.~{Chiotakis}, S.~{Pinneterre}, and M.~{Paolino}, ``vfpgamanager: A
  hardware-software framework for optimal fpga resources exploitation in
  network function virtualization,'' in \emph{Proc. of EuCNC}, June 2019, pp.
  47--51.

\bibitem{Paone14OCL}
E.~{Paone} \emph{et~al.}, ``Evaluating orthogonality between application
  auto-tuning and run-time resource management for adaptive {OpenCL}
  applications,'' in \emph{Proc. of ASAP}, 2014, pp. 161--168.

\bibitem{Gadioli18Socrates}
D.~{Gadioli} \emph{et~al.}, ``{SOCRATES} — a seamless online compiler and
  system runtime autotuning framework for energy-aware applications,'' in
  \emph{Proc. of DATE}, 2018, pp. 1143--1146.

\bibitem{khasanov_date20}
R.~Khasanov and J.~Castrillon, ``Energy-efficient runtime resource management
  for adaptable multi-application mapping,'' in \emph{Proc. of DATE}, Mar.
  2020, pp. 909--914.

\bibitem{vitali19}
E.~{Vitali} \emph{et~al.}, ``An efficient monte carlo-based probabilistic
  time-dependent routing calculation targeting a server-side car navigation
  system,'' \emph{IEEE Transactions on Emerging Topics in Computing}, 2019.

\bibitem{did2019}
D.~{Diamantopoulos} and C.~{Hagleitner}, ``Helmgemm: Managing gpus and fpgas
  for transprecision gemm workloads in containerized environments,'' in
  \emph{Proc. of ASAP)}, vol. 2160-052X, 2019, pp. 71--74.

\bibitem{lagasio2019predictive}
M.~Lagasio \emph{et~al.}, ``Predictive capability of a high-resolution
  hydrometeorological forecasting framework coupling wrf cycling 3dvar and
  continuum,'' \emph{Journal of Hydrometeorology}, vol.~20, no.~7, 2019.

\bibitem{lagasio2019synergistic}
M.~Lagasio \emph{et~al.}, ``A synergistic use of a high-resolution numerical
  weather prediction model and high-resolution earth observation products to
  improve precipitation forecast,'' \emph{Remote Sensing}, vol.~11, no.~20,
  2019.

\bibitem{Golasowski20}
M.~Golasowski \emph{et~al.}, ``Alternative paths reordering using probabilistic
  time-dependent routing,'' \emph{Advances in Intelligent Systems and
  Computing}, vol. 1036, pp. 235--246, 2020.

\bibitem{ptosek18}
V.~Ptošek \emph{et~al.}, ``Real time traffic simulator for self-adaptive
  navigation system validation,'' in \emph{Proc. of EMSS}, 2018, pp. 274--283.

\end{thebibliography}

\end{document}